\pdfoutput=1
\documentclass[12pt]{article}
\usepackage{graphics, color}
\usepackage{graphicx}
\usepackage{amssymb,mathtools}

\usepackage{hyperref}

\numberwithin{equation}{section}

\def\gsim{\, \rlap{$>$}{\lower 1.1ex\hbox{$\sim$}}\,}
\def\lsim{\, \rlap{$<$}{\lower 1.1ex\hbox{$\sim$}}\,}

\newcommand{\be}{\begin{equation}}
\newcommand{\ee}{\end{equation}}
\newcommand{\bea}{\begin{eqnarray}}
\newcommand{\eea}{\end{eqnarray}}
\newcommand{\vol}{{\cal V}}

\newcommand{\ba}{\begin{eqnarray}}
\newcommand{\ea}{\end{eqnarray}}
\newcommand{\nn}{\nonumber}

\begin{document}

\begin{titlepage}

\setcounter{page}{1} \baselineskip=15.5pt \thispagestyle{empty}

\vbox{\baselineskip14pt
%\hbox{hep-th/0000000}
}
{~~~~~~~~~~~~~~~~~~~~~~~~~~~~~~~~~~~~
~~~~~~~~~~~~~~~~~~~~~~~~~~~~ \footnotesize{SU/ITP-14/13, SLAC-PUB-15962, DESY-14-078}} \date{}

\bigskip\

\vspace{.5cm}
\begin{center}
{\fontsize{19}{36}\selectfont  \sc
The Powers of Monodromy\\
\vspace{4mm}
}
\end{center}

\begin{center}
{\fontsize{13}{30}\selectfont Liam McAllister,$^{1}$ Eva Silverstein,$^{2,3,4}$
Alexander Westphal,$^{5}$ and Timm Wrase$^{2}$}
\end{center}

\begin{center}
\vskip 8pt
\textsl{$^1$ Department of Physics, Cornell University, Ithaca, NY 14853, USA}

\vskip 7pt
\textsl{$^2$ Stanford Institute for Theoretical Physics, Stanford University, Stanford, CA 94305, USA}

\vskip 7pt
\textsl{$^3$ SLAC National Accelerator Laboratory, 2575 Sand Hill Rd., Menlo Park, CA 94025, USA}

\vskip 7pt
\textsl{$^4$ Kavli Institute for Particle Astrophysics and Cosmology, Stanford, CA 94305, USA}

\vskip 7pt
\textsl{$^5$Deutsches Elektronen-Synchrotron DESY, Theory Group, D-22603 Hamburg, Germany}

\end{center}

\vspace{0.2cm}
\hrule \vspace{0.1cm}
{ \noindent \textbf{Abstract} \\[0.2cm]
\noindent Flux couplings to string theory axions yield
super-Planckian field ranges along
which the axion potential energy grows.
At the same time, other aspects of the physics
remain essentially unchanged along these large displacements, respecting a discrete shift symmetry with a sub-Planckian
period.
After a general overview of this monodromy effect and its application to large-field inflation,
we present new classes of specific models of monodromy inflation, with monomial potentials $\mu^{4-p}\phi^p$.
A key simplification in these models is that the inflaton potential energy plays a leading role in moduli stabilization during inflation.
The resulting inflaton-dependent shifts in the moduli fields lead to an effective flattening of the inflaton potential, i.e.~a reduction of the exponent from a fiducial value $p_0$ to $p<p_0$.
We focus on examples arising in compactifications of type IIB string theory on products of tori or Riemann surfaces,  where the inflaton descends from the NS-NS
two-form potential $B_2$, with monodromy induced by a coupling to the R-R field strength $F_1$.
In this setting we exhibit models with $p=2/3,4/3,2,$ and $3$, corresponding to  predictions for the tensor-to-scalar ratio of $r\approx 0.04, 0.09, 0.13,$ and $0.2$, respectively.
Using mirror symmetry, we also motivate a second class of examples with the role of the axions played by the real parts of complex structure moduli, with fluxes inducing monodromy.}

\vspace{0.3cm}
 \hrule

%\vspace{0.6cm}
\end{titlepage}

\tableofcontents

\newpage

\baselineskip = 16pt

\section{Motivation and Overview}

\noindent

Cosmological observables provide a window into very early times in our universe,
offering a unique set of probes of high-energy physics.  In particular, in the context of
inflation \cite{inflation}, an observation of a  tensor-to-scalar ratio $r \gtrsim 10^{-2}$  implies an unprecedented connection between empirical
observations and quantum gravity, for two reasons: it provides a measurement of the
quantum mechanical variance of the tensor modes of the metric \cite{earlyperts}\cite{Snowmass},
and it indicates a super-Planckian field excursion \cite{Lyth}.
An impressive variety of observational  efforts are  approaching the sensitivity  required  to detect
$r$ in this range \cite{Bmodeexp}, with a recent report of a detection of B-mode polarization \cite{BICEP}\ that may contain a signal of primordial origin corresponding to inflationary tensor modes \cite{Bmodetens}, depending on the outcome of important foreground measurements generalizing \cite{Planckdust}.

The inflationary energy density in large-field inflation is sub-Planckian --- albeit relatively high, $\sim (10^{16}\, {\rm{GeV}})^4$ --- so that the process can be described, and was originally discovered theoretically, in the context of low-energy quantum field theory coupled to gravity.  But the large field range implies sensitivity to
an infinite sequence of {\it dangerously irrelevant} Planck-suppressed operators.  Low-energy field theory models of large-field inflation can be radiatively stable, and natural in the sense of Wilsonian renormalization, by virtue of an approximate shift symmetry.   However,  imposing such a symmetry, even at the classical level, amounts to making a strong assumption about the ultraviolet (UV)  completion of the inflationary effective theory.
It would be much more satisfying --- and in our view it is necessary --- to understand how  the structures required for large-field inflation emerge from a complete theory of quantum gravity.

It is tempting to belabor the motivation for modeling inflation in string theory by drawing examples from other subjects, such as condensed matter physics, illustrating the importance of the `ultraviolet'-complete treatment in the presence of sensitivity to irrelevant operators in the effective theory.\footnote{We thank S. Hartnoll and S. Kachru for discussions.}
Even though one can model certain low-energy phenomena such as superconductivity using a continuum field theory,  knowledge of the microphysics is required to understand very basic aspects of the problem, such as the transition temperatures available in real materials.  For example,  in applying BCS theory to metals one needs to recognize that the attractive interaction yielding Cooper pairs arises from phonons.  Low-energy theory alone would suggest a much wider variety of transition temperatures than is observed in nature, a discrepancy that may be due to constraints  from the UV completion of the system.
More generally,
important
aspects of the physics (such as transport) can be described by an irrelevant operator, and thus be sensitive to aspects of the  UV theory (such as the breakdown of translation invariance due to the lattice).   Moreover, certain effects, such as the melting of a solid, are strongly UV sensitive.  Although one can model a wide variety of behaviors in low-energy field theory, it would
be a mistake to work purely in a low-energy effective description, ignoring the structure and constraints implied by the ultraviolet theory.

Of course, the major difference in the present case is that we do not know the correct theory of quantum gravity, whereas in the condensed matter analogue the relevant short distance theory is standard.   But that is a logically independent point, and does not diminish the importance of obtaining large-field inflation from a complete theory of quantum gravity.

Without  detailed knowledge of the UV completion of gravity, one might worry that as the inflaton moves over a Planckian range in field space --- or more generally a range $M_{UV}\le M_P$, where $M_{UV}$ characterizes the scale of new physics involved in quantum gravity --- new degrees of freedom could become important in the dynamics.  These new degrees of freedom could be different in different parts of the long field excursion, and lead to independent contributions to the potential that strongly violate the slow roll conditions.
Note that this is already an important question at the {\it{classical}} level: although a shift-symmetric model  can be radiatively stable, and hence internally consistent from the low-energy effective field theory point of view, whether a  given shift symmetry  admits an ultraviolet completion in quantum gravity  requires careful examination.

String theory is a very promising candidate  theory of quantum gravity, with many concrete successes in the arena of thought experiments and mathematical and physical consistency checks.   The strong evidence for its consistency includes precise black hole entropy counts, the AdS/CFT correspondence,  the perturbative finiteness of the theory,  its role in resolving singularities,  the intricate duality relations  that make sense of various strong coupling and high-curvature limits, and the capacity of its landscape of vacua to accommodate the small cosmological constant (as a selection effect).
Despite the astronomical number of solutions of string theory, the mathematical structure of the theory remains highly constrained.

\begin{figure}[htbp]
\begin{center}
\includegraphics[width=15cm]{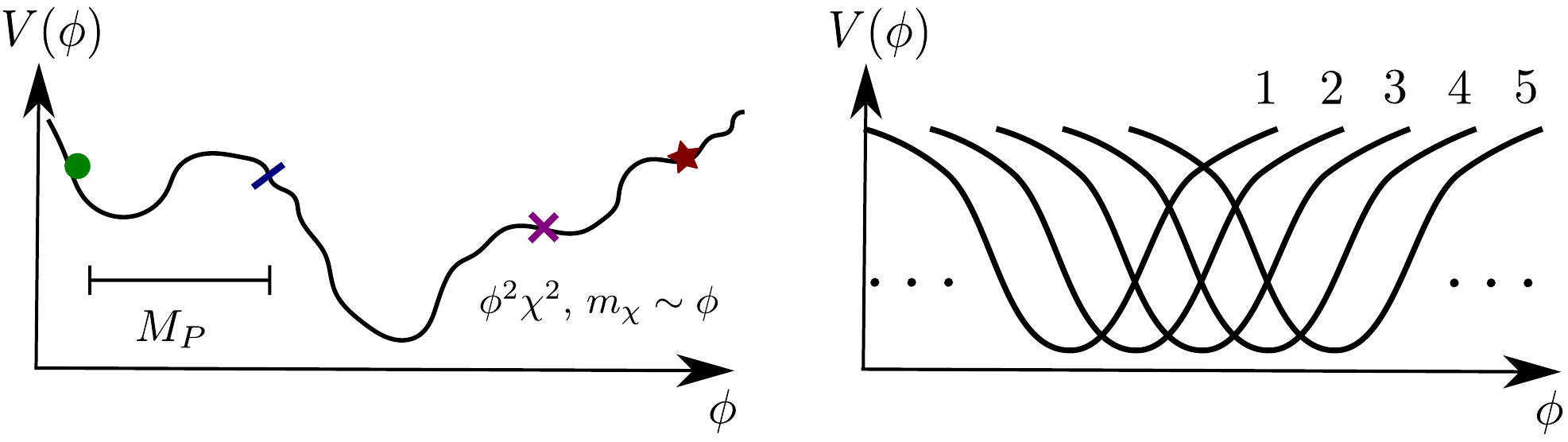}
\end{center}
\caption{On the left, a sketch of a large field range with new effects --- such as altered couplings or new light states --- appearing after each displacement of order $\sim M_P$,  parameterizing our ignorance of quantum gravity.  Such features could arise both at the classical and the quantum level.  On the right, the structure of the potential along axion directions (and their various duals) in string theory.  The whole structure has a sub-Planckian period $f$, but on each branch the field can reach a
large field range.   The potential energy grows  with each cycle around the underlying period $f$, while
other conditions --- such as the spectrum of branes wrapping the cycles threaded by the higher dimensional potential fields yielding axions --- remain the same each time around.  The result  is a radiatively-stable potential as in chaotic inflation  with a monomial potential. \label{strawmanargument}
}
\end{figure}

The  microphysical
structure of string theory provides a rather simple and general mechanism for large-field inflation \cite{Andreichaotic}, {\it{monodromy}} \cite{monodromyI}\cite{monodromyII}\cite{ignoble}, in which
an underlying periodicity of the theory ensures that as the inflaton field traverses many cycles with sub-Planckian period $2\pi f\ll M_P$, the  potential energy increases over each cycle but much of the remaining physics essentially repeats itself (see figure \ref{strawmanargument}).

In this work, we will begin a more systematic analysis of the monodromy effect in string theory and its application to inflationary cosmology.  As we will review in detail below, the couplings of
axions to fluxes exhibit  monodromy in a robust way.
In the presence of sufficiently generic fluxes or brane configurations, the field ranges of axion fields  (and their duals)
extend to super-Planckian values,  but the underlying sub-Planckian
periodicity governs much of the physics along the trajectory \cite{monodromyI}\cite{monodromyII}\cite{ignoble}\ including significant sectors of the spectrum of particles and branes.
Starting from this general framework, we will provide a new class of specific examples of monodromy inflation in string theory, with a variety of values of the tensor-to-scalar ratio, including some with significantly larger values of $r$ than in previously studied realizations of
monodromy in string theory.
The  cause of the large values of $r$ is very simple:  in our examples, inflation is driven by  potential  energy terms that involve  moderately high powers of the axion field.

Although the monodromy  phenomenon in itself is quite simple, there are substantial complications involved in modeling inflation explicitly in string theory:   a primary problem  is the stabilization of the many  moduli fields that arise upon compactification.
In the  new models presented here, moduli stabilization is simpler in one respect than in previous realizations:  the inflationary energy itself plays a leading role in determining the vevs of some of the moduli,  which shift adiabatically as inflation proceeds.   This flexibility allows for successful moduli stabilization  even when the inflationary energy is large enough to invalidate a more rigid stabilization scenario.
Moreover,  the shifting of the moduli alters the form of the inflaton potential, while not disrupting inflation.
With the adjustment of the moduli fields consistently taken into account, we find
potentials that take the form of a sinusoidally modulated power law,
\be\label{Vform}
V(\phi)\approx \mu^{4-p}\phi^p + \Lambda^4 \sin\left( \frac{\phi}{f}\right) \, ,
\ee  over the relevant range of the inflaton field $\phi$.
Previous work had exhibited a concrete example with $p=1$ \cite{monodromyII}, and motivated a variety of others with various powers $p$ \cite{monodromyII}\cite{ignoble}\cite{flattening} \cite{AlbionRoberts}\cite{unwinding}\cite{Shlaer:2012by}\cite{gurari}\cite{recentmonodromy}.  In the present work, we will derive a broader range of powers $p$ from a wider variety of flux-induced axion interactions.\footnote{Other interesting recent work on axion inflation in string theory without incorporating monodromy appears in \cite{recentnonmonodromy}, building on earlier works such as \cite{natural}\cite{Kim:2004rp}\cite{Nflation}\cite{Grimm}.}

The fields of primary interest here
arise from  the internal components of higher dimensional potential fields (of various ranks), which are generalizations of the vector potential $A$ of electromagnetism.
The axions descend from rank-$p$ potential fields $A_p$ as
$A_p={a} \omega_p$, where $\omega_p$ is a $p$-form in the cohomology of the internal space.  For example, a one-form $A$
integrated around a circle $S^1$ in the extra dimensions gives an axion, ${a}=\int_{S^1} A$.  There is a rich set of gauge invariant terms in the low-energy Lagrangian of compactified string theory  that exhibit direct dependence on these potential fields and on their field strengths; these generalize Stueckelberg terms $(\partial C+A)^2$
that are gauge-invariant under $A\to A+\partial\Lambda, C\to C-\Lambda$.

In \S\ref{sec:generalities}, we will give an overview of the monodromy arising from the
couplings between fluxes and axion fields in string theory, paying particular attention to the contributions  relevant for
our new examples.  The couplings to fluxes  of an axion field $a$ produce a potential that at large $a$ takes the  form
$V \sim f(\chi)\times a^{p_0}$
where $\chi$ represents the moduli fields, including massive Kaluza-Klein scalars, and $p_0$ is an integer.
Backreaction  of the inflationary energy on the  moduli $\chi$,
which adjust in an energetically favorable way, can change the shape of the potential at large field values. Previous work focused on examples with $p_0=2$; in one canonical class of examples this fiducial power is `flattened' to the linear potential $V\propto  a^{1}$ originally studied in \cite{monodromyII}, as can be seen explicitly using the gravity-side description of the corresponding brane system \cite{flattening}.  In the present work, we will include examples in which higher-rank wedge products of the rank-two potential field $B$ sourced by fundamental strings lead to higher fiducial powers $p_0$.   In particular,
we find examples with $p_0=4$.  We will also comment on dual cases, including axion-like components of complex structure moduli of Calabi-Yau and Riemann surface compactifications.  After an instructive warmup example in \S\ref{sec:UVB}\ exhibiting flattening along a complex structure direction, we will present string compactifications that realize monodromy inflation in \S4.  In these examples, the flattening effect leads to
a
variety of final powers including $p=3,2,4/3$ and $2/3$.
In \S\ref{sec:UVII}, we will make comments on the monodromies in complex structure moduli space from flux potentials in Calabi-Yau and Riemann surface compactifications.

\section{Flux Couplings and
Monodromy}
\label{sec:generalities}

In this section we explain the origin of terms in the effective action that have a monodromy-induced potential growing as an integer power
$p_0\ge 2$
of an axion field, while also depending on moduli fields coming from the internal metric and the dynamical string coupling.   In \S\ref{sec:UVB},
we will show explicitly how these terms can lead to a variety of power law potentials $V\propto \phi^{p\le p_0}$, with the final power $p$ shifted down from $p_0$ via adjustments of heavy moduli.

\subsection{Axions from the two-form potential $B$}\label{sec:setupB}

Perturbative string theory contains a two-form potential field
$B=B_{MN}dx^M\wedge dx^N$ that is directly analogous to the usual vector potential $A=A_M dx^M$ of electromagnetism.\footnote{An exception is the type I string, in which closed strings are unstable to breaking into open strings, but this theory contains a two-form potential sourced by D1-branes.}  In particular,
$B$ is sourced by fundamental strings just as
the usual vector potential is sourced by charged particles.   There is a gauge invariance in the theory under which $B\to B+d\Lambda_1$, with $\Lambda_1$ a one-form, analogous to the gauge invariance under $A\to A+d\Lambda_0$ in electromagnetism.
Similarly, there are other potential fields denoted $C_{p+1}$ sourced by $p$-dimensional extended objects (D$p$-branes) \cite{Polchinski}.

In electromagnetism, the action contains the gauge-invariant terms
\be
S_{EM}=\int d^4 x \sqrt{-g}\left\{ F_{MN} F^{MN} - \rho^2 (A_M+\partial_M C)^2+\dots\right\}\, ,
\ee
where under the gauge transformation $A_M\to A+\partial_M\Lambda_0$, the field $C$ transforms as $C\to C-\Lambda_0$.  The first term is the Maxwell action, written in terms of the field strength $F=dA$.  The second term, known as a Stueckelberg term, can arise from spontaneous symmetry breaking, with $\rho$ the vacuum expectation value of a charged field.\footnote{In ordinary electrodynamics the symmetry is of course unbroken in vacuum, but $\rho\ne 0$ arises in a superconductor from the condensation of the Cooper pair field.}

In type II string theory, one finds generalizations of these Maxwell and Stueckelberg terms, with the
gauge transformation $B\to B+d\Lambda_1$ accompanied by appropriate shifts of the $C_p$ fields.
Although we will focus on specific examples in type IIB string theory below, let us start by considering the relevant terms arising in $D=10$ type IIA string theory.  There we have potential fields $C_p$ with odd $p$, and it is useful to define the following generalized field strengths that respect all the gauge symmetries of the theory:
\bea\label{genfld}
H_3 &=& dB\, ,\nn\\
F_0 &=& Q_0\, , \nn\\
\tilde F_2 &=& d C_1+ F_0 B\, ,\nn\\
\tilde F_4 &=& dC_3+C_1\wedge H_3 + \frac{1}{2} F_0 B\wedge B\, ,
\eea
where $Q_0$ is an integer.  These are gauge-invariant, with the transformation $B\to B+d\Lambda_1$ extended to a combined transformation
\bea\label{gaugetransf}
\delta B &=& d\Lambda_1\, , \nn\\
\delta C_1 &=& -F_0\Lambda_1\, , \nn\\
\delta C_3 &=& -F_0 \Lambda_1\wedge B\, .
\eea
The effective action starting from a total dimensionality $D=10$ contains
terms proportional to\footnote{Similar comments apply in the more generic cases with $D>10$ \cite{newaccel}.}
\be
-\frac{1}{{\alpha'}^4}\int d^{10} x\sqrt{-G}\Bigl\{\frac{1}{g_s^2} |H_3|^2+ \sum_p|\tilde F_p|^2\Bigr\}\, .
\ee
Upon dimensional reduction to four dimensions, these terms introduce a direct dependence of the potential energy on the axion fields
\be\label{bax}
b^i \equiv \int_{\Sigma_2^i} B
\ee
obtained by integrating the potential field $B$ over nontrivial 2-cycles $\Sigma_2^i$ in the  compactification manifold ${\cal M}$.
Another feature we need to take into account is that the fluxes $Q_2^i=\int_{\Sigma_2^i} dC_1, Q_4=\int_{\Sigma_4^i} dC_3$, and  $N_3=\int_{\Sigma_3^a} H_3$ (with  the index $i$  running over topologically distinct even-dimensional cycles, and $a$ similarly indexing three-cycles) are quantized, as is $Q_0=F_0$.

Let us focus on the $B$-dependent terms, and for simplicity work on the branch of the potential where $Q_2=Q_4=0$ (also setting to zero the flux $dC_3$ along the noncompact four dimensional spacetime, or equivalently the dual 6-form flux $Q_6\equiv \int_{{\cal M}}\star_{10} F_4=\int_{{\cal M}}F_6$).
In the models in \S\ref{sec:UVB}, we will incorporate the analogue in type IIB string theory of these additional fluxes, which will yield interesting behavior in some cases, but for now we will focus on the leading contributions to the potential at large field range.
Given this, we have an action of the schematic form\footnote{See e.g. equation 12.1.25 of \cite{Polchinski}.   However, we caution the reader that we follow the sign conventions of \cite{massiveII}, not those of \cite{Polchinski}.}
\be\label{mzeroB}
-\frac{1}{{\alpha'}^4}\int d^{10} x \sqrt{-G}\left\{\frac{1}{g_s^2}|H_3|^2 + |Q_0 B|^2 + |Q_0 B\wedge B|^2  + \gamma_4 g_s^2 |Q_0 B\wedge B|^4+\dots \right\}\, .
\ee
Here in the last term and the ellipses we have allowed for corrections that could be read off from the tree-level four-point and higher-point functions ($\gamma_4$ being an order 1 number).  We have also set to zero the contribution from $|\tilde F_6|^2=|C_3\wedge H_3+Q_0 B\wedge B\wedge B/6|^2$, having in mind situations where $H_3$ flux
is present in order to contribute to moduli stabilization, and $C_3$ minimizes the $|\tilde F_6|^2$ term at zero.  More generally, there should be interesting configurations in which $C_3\wedge H_3 \neq -Q_0 B\wedge B\wedge B/6$ at the $C_3$ minimum, or configurations in which $C_3$ and $B$ evolve together, in which cases this term is relevant.

The field strengths of R-R terms come with a factor of $g_s$, so higher-dimension operators  involving higher powers of generalized field strengths $\tilde F_p$ --- even those from string tree diagrams --- appear with a relative factor of $Q_0^2g_s^2$, and are thus suppressed at small string coupling.  This is in the standard frame we will use exclusively here, where gauge transformation and flux quantization conditions are most simply expressed.

In fact, there is generically an additional suppression factor at large radius.  We will shortly consider generalizations that arise upon dimensional reduction or T-duality, where $F_0$ is replaced by higher-form fluxes $F_n$.  In those cases, the suppression is even stronger, with each power of $|\tilde F|^2$ coming with a factor of $g_s^2Q_n^2/L^{2n}$, where $L$ is the size in string units of the cycle threaded by the $F_n$ flux.

Below, we will consider specific examples in type IIB string theory with effective $|F_1 B|^2 +|F_1\wedge B\wedge B|^2$ interactions.  These follow from T-duality of (\ref{mzeroB}) upon reduction of the IIA theory on a circle as explained in detail in \cite{massiveII}.  At first glance, this is not manifest from the  generalized fluxes that appear in the type IIB equations of motion in ten dimensions:
\bea\label{FtildeB}
H_3 &=& dB\, ,\nn\\
F_1 &=& dC_0\, , \nn\\
\tilde F_3 &=& dC_2-C_0 H_3\, ,   \nn\\
\tilde F_5  &=&  dC_4-\frac{1}{2} C_2\wedge H_3 + \frac{1}{2}B\wedge dC_2\, .
\eea
In $\tilde F_5$ we do not find an $F_1\wedge B\wedge B$ term  by working directly in the ten-dimensional theory.
However, T-duality on a circle, including the duality between D7-branes and D8-branes, requires this coupling to be present upon dimensional reduction.  This indeed works out precisely \cite{massiveII}.
Specifically, consider  reducing   ten-dimensional  type IIB theory on a circle (along the $x^9$ direction, $x^9 \cong x^9+2\pi$), with
\bea\label{IIBfields}
C_0 &=& x^9 Q_0 + {\cal C}_0\, , \nn\\
C_2 &=& x^9 Q_0 B + {\cal C}_2\, ,
\eea where ${\cal C}_p$ are fluctuations of the potential fields about the background.
Substituting (\ref{IIBfields}) into (\ref{FtildeB}), we find
an effective $F_1\wedge B\wedge B$ contribution to $\tilde F_5$, and an effective $F_1\wedge B$ term in $\tilde F_3$.
In the four-dimensional effective theory, there are many contributions of this kind, leading to axion potentials of the schematic form
\be\label{schematicpot}
f(\chi,\dots) ~ \frac{(Q^{(n)} a^{n}+Q^{(n-1)} a^{n-1}+\dots + Q^{(0)})^2}{L^{2 n^{\prime}}}+\dots \sim \tilde f(\chi,\dots) ~ a^{p_0}  ~~~
{\rm for}~a \gg 1\,,
\ee
where we have  denoted the axion field by $a$, $n=p_0/2$ is a  positive
integer, and ``$\chi,\dots$" refers to the moduli fields $\chi$, as well as additional scalar fields,
whose important effects we will analyze below.  The value of $n$ depends on the ranks of the fluxes and potential fields that descend to the $Q^{(i)}$ and to $a$, respectively; we will discuss specific examples in the following section.

We see immediately from (\ref{schematicpot}) the branch structure of the monodromy-unwound potential:  for a fixed value of the flux quantum number $Q^{(0)}$ here, the potential is a growing function of $a$, which has an unbounded field range (up to the point where the potential energy density  becomes so large that the low-energy description breaks down).  The whole structure, on the other hand, is periodic under shifts of $a$ by an integer, as this can be absorbed by an appropriate shift of the flux quantum numbers.
That is, there is an identical branch of the potential for each value of $Q^{(0)}$, as in figure \ref{strawmanargument}.  Similarly, the {\it spectrum} of particles and higher dimensional branes
that couple to $a$ is periodic\footnote{The subsectors of the spectrum coming from wrapped branes as described in the text are periodic, while other sectors of metastable states can be affected by the  monodromy along the branch of the potential on which the system inflates.
See \cite{monodromyII} for explicit examples of both classes, with the latter case arising from modes living on a spacetime-filling brane.  These latter sectors are the closest the system comes to the emergence of light states at large field \cite{oogurivafa}, albeit not via an approach to a weak-coupling or large-radius limit of moduli space.} under $a \to a+1$.  For example, one gets D-strings from wrapping a D3-brane on a two-cycle threaded by $B$.  A given wrapped D-brane gives a string in four dimensions with a tension that grows with the axion $b=\int B$, but the {\it set} of wrapped D-branes is invariant under shifts of $b$ by an integer.
This provides a reasonably clean answer to the question of controlled large field ranges in quantum gravity.

As we will analyze in detailed examples below, the moduli-dependence in the potential energy has important effects on the inflationary dynamics.  If the moduli (and other massive degrees of freedom, such as Kaluza-Klein modes, included in the ellipses in $\chi,\dots$) are stabilized very rigidly, the inflationary potential could end up behaving like $a^{p_0}$.
That requires the inflationary potential to be subdominant to the leading terms stabilizing the moduli.  More generally, as we increase the vev of the axion field $a$, the other fields will adjust  in response to the potential energy carried by the term (\ref{schematicpot}).    This can be an important effect even for fields more massive than the Hubble scale during inflation, as was first pointed out, and explored in various cases with $p_0=2$, in \cite{flattening}.  As an example, the linear potential of \cite{monodromyII} arises in a simple way as a flattening effect from $p_0=2$ to $p=1$.

The couplings we have reviewed above can produce examples in which the fiducial power $p_0$ is either 2 (e.g. from $|dC_2\wedge B|^2$) or 4 (e.g. from $|F_n \wedge B\wedge B|^2$, $n=0$ or $1$).
As we have seen, some of these couplings are manifest from a simple dimensional reduction of the terms in the higher-dimensional type II string theory action \cite{Polchinski}.  Other such terms come from appropriate field configurations, as in (\ref{IIBfields}), that can arise in the reduction of the higher-dimensional theory.  Some of these two types of terms, and many others, can be related to each other by duality symmetries.

A rich set of string dualities also relate the $B$ potential fields we have focused on here to the R-R  potentials $C_p$, and also to other scalar fields such as the real parts of complex structure moduli (related to $B$ via T-duality or its generalizations like mirror symmetry).  We will comment further on the latter case below.
It would take us too far afield to enumerate all the possible fields and terms, but it is clear that the monodromy effect is ubiquitous --- to avoid it requires turning off fluxes and/or choosing internal manifolds with special topology.

One final comment on genericity is worth making here:
in this work we will consider string theory in $D=10$ dimensions, but $D>10$ limits of string theory also exhibit axion-flux couplings with a similar structure, including
important couplings to other scalar fields.  In
such cases,
one might find even higher fiducial powers $p_0$, which when combined with heavy-field adjustments could lead to a larger range of potentials,
which  could
be analyzed explicitly as in \cite{newaccel}.  In fact, in $D>10$ the spectrum is dominated by axions from R-R potential fields, whose number grows as $2^D$.  In $D=10$, axions and their duals are an order-one fraction of the scalar fields, and hence already rather generic as candidate inflatons.

Before moving to specific examples in the next section, let us continue to study the general structure of the potential and how it behaves at large field range, in the regime where the canonically normalized field takes super-Planckian values.
As already discussed, the axions $b_i$ from the two-form potential $B$ arise from cohomology elements in the internal manifold,
\be\label{Bexp}
B=\sum b_i\omega_{2}^i\, ,
\ee
where $\omega_{2}^i$ are nontrivial two-forms.  The relation between the axions $b_i$ and the canonical field depends on the geometry, and specifically on $\int \omega\wedge \star\omega$.
For simplicity let us first consider a situation in which
all length scales are comparable, of order $L$ in string units,  and there is no significant warping.
From the kinetic term $\int |H_3|^2$ for $B$ we get the canonical fields
\be\label{can}
\phi_i\sim f b_i \sim b_i \frac{M_P}{L^2}\, ,
\ee
and an effective action of the form
\be\label{fourd}
\int d^4 x\sqrt{-g} \left\{ \sum_i g^{00}\dot\phi_i^2 - V(\phi_i; \sigma_I)\right\}\, ,
\ee
where $\sigma_I$ denotes moduli fields and other degrees of freedom such as those related to the internal spatial profiles of the fields (i.e., Kaluza-Klein modes).
With multiple fields there  is the possibility of kinetic mixing, as we  will discuss below.

With the above approximations we arrive at a potential of the form
\ba\label{Vscales}
V&\sim& M_P^4  \frac{g_s^4}{L^6}\frac{Q_n^2}{L^{2n}} \left(\frac{b^2}{L^4}+\frac{b^4}{L^8}+{\cal O}\left(\frac{g_s^2 Q_n^2}{L^{2n}} \frac{b^8}{L^{16}}\right)\right)\nonumber\\
&&\\
& \sim & M_P^4  \frac{g_s^4}{L^6}\frac{Q_n^2}{L^{2n}} \left(\frac{\phi^2}{M_P^2}+\frac{\phi^4}{M_P^4}+{\cal O}\left(\frac{g_s^2 Q_n^2}{L^{2n}} \frac{\phi^8}{M_P^8}\right)\right)\, .\nonumber
\ea
Here we have assumed  that the configuration of fluxes and
axion(s) $b_i\sim b$ is sufficiently generic so that
$F_n\wedge B\wedge B \neq 0$,  leading to  a potential term quartic in $b$.  In other special cases, e.g. in type IIB theory without the background field configuration in (\ref{IIBfields}), the quartic term may be absent, leading to a quadratic fiducial potential.

The expression (\ref{Vscales}) has two main implications for our purposes.  First,
at least in this one-scale situation, the quartic term dominates in the super-Planckian regime $\phi\gg M_P$.\footnote{Although we  have illustrated  this point  in a system with a single length scale, the result is more general.  In fact, in configurations with multiple length scales, as described around (3.29) of \cite{flattening}, the higher powers of $\phi$ can dominate even for $\phi < M_P$.}
Second,  the higher-dimension operators coming from higher powers of $|\tilde F_p|^2$ are negligible
as long as  $g_s^2/L^{2n} \ll 1$.  In (\ref{Vscales}) we took into account that the largest power of $\phi/M_P$ dominates in these terms suppressed by $g_s^2/L^{2n}$.
This requires moderately large radius and small string coupling.  From now on, we will drop higher-dimension terms
for this reason.

In the next section, we will incorporate backreaction on the moduli fields $\chi$, finding specific examples in which the fiducial power $p_0$ is shifted to various powers $p\le p_0$ depending on the interplay of the leading large-field inflationary potential term and the other terms in the moduli potential,
\be\label{Vphipowers}
V\sim f(\chi)~\phi^{p_0}+V_0(\chi) ~~ \to ~~ V(\phi)\approx \mu^{4-p}\phi^p +\Lambda^4 \sin\left( \frac{\phi}{f}\right)\, .
\ee
In the last term we allowed for periodic contributions, which are suppressed at large radius.

\subsection{Radiative stability}

For completeness, let us briefly review radiative stability in large-field inflation.  Chaotic inflation with a monomial potential \cite{Andreichaotic}, including generalizations to non-integer powers $p$ via  monodromy, as in (\ref{Vphipowers}), is radiatively stable.  The couplings intrinsic to the power law potential (expanding in field perturbations $\delta\phi$) become smaller at large field range, and gravitational interactions are also suppressed \cite{smolin}.
In effect, as long as the inflationary scalar potential constitutes the leading source of shift symmetry breaking, the model is technically natural in the sense of 't Hooft (and can be fully natural in the sense of Wilson given dynamically small scales).
On the other hand, establishing that the approximate  shift symmetry  encoded in (\ref{Vphipowers})  can arise in a consistent quantum gravity theory requires  careful consideration of the ultraviolet completion: in the context of string compactifications,  it is necessary in particular  to  verify that the symmetry survives stabilization of all moduli.   We  check this in explicit examples in \S\ref{iibRiem} below.

\subsection{Dual axions}\label{sec:RiemannMirrorAxions}

Before moving on to our main examples, we briefly mention other fields, related to the $B$ field by dualities, that undergo monodromy in the presence of appropriate fluxes.\footnote{Our discussion here is not exhaustive: other examples  include configurations with moving branes \cite{monodromyI}\cite{unwinding}\cite{gurari}, as well as some of the  scenarios in \cite{recentmonodromy}; some of these may also be understood via dualities.}

First,  as noted above, string dualities relate the NS-NS two-form $B$ to the R-R $p$-forms $C_p$.
We will not analyze such examples here,  but characterizing them would be a large part of a systematic analysis of the monodromy mechanism.  As an example, in one of the original models \cite{monodromyII}\ of axion monodromy,  the inflaton is an  axion descending from $C_2$.

Next, complex structure moduli of certain special compactification manifolds, such as Calabi-Yau manifolds and Riemann surfaces, contain components that behave like axions, i.e.~fields that are periodic in the absence of monodromy-inducing sources.
These are sometimes related by string dualities to the axions descending from higher dimensional components of the various gauge potentials. The motion of 7-branes is on the same footing in some sense, as 7-brane position moduli arise from complex structure moduli in F-theory.\footnote{See \cite{Bizet:2014uua} for a recent paper that determines monodromies on CY$_4$ manifolds.}

For the simplest example, consider string theory on a two-dimensional torus $T^2$.  The complex structure modulus $\tau$ describes the same geometry if
we shift $\tau\to\tau+1$.  This complex structure modulus $\tau$ is related by a T-duality (or mirror symmetry)  transformation to a
modulus $\rho=b+i\, \mathrm{Vol}(T^2)$, where $b=\int_{T^2}B$ and $\mathrm{Vol}(T^2)$ is the volume of the two-torus.
The underlying periodicity of $b$ is mirror to the transformation $\tau\to\tau+1$ of the real part of the complex modulus $\tau$.
Flux threading one cycle of this $T^2$ removes the periodicity under $\tau\to \tau+1$, inducing monodromy.
For Riemann surfaces with genus $h>1$, a similar effect arises,
at least in a limit of complex structure where the surface nearly factorizes into $h$ tori separated by thin necks.

In Calabi-Yau manifolds, mirror symmetry relates the axions from $B$ to components of the complex structure moduli.  The $B$ fields have an underlying periodicity, realized as a set of $\theta$ angles in a gauged linear sigma model \cite{GLSM}\ treatment of Calabi-Yau manifolds.  The complex structure moduli exhibit a corresponding monodromy:  the periods, and hence the fluxes, do not return to themselves after going around special points in the moduli space.  About large complex structure, for example, there is a monodromy group $\mathbb{Z}$ as in the $T^2$ case just discussed.

The flux stabilization potential for complex structure moduli of Calabi-Yau manifolds \cite{GKP}\ as well as of Riemann surfaces \cite{Saltman}\ contains a sextic potential for these complex-structure dual-axions, at fixed values of the remaining moduli.  In some cases, such as the examples in \cite{Saltman}, this flux potential for the complex structure moduli also provides a leading contribution to the stabilizing potential for the volume and the string coupling.

In \S\ref{sec:UVII} we will remark briefly on the complex structure analogue (roughly the mirror) of the examples based on $B$ axions that appear in \S\ref{sec:UVB} and \S\ref{iibRiem}.
It would be worthwhile to analyze more systematically the possibility of complex structure monodromies for inflation.

\section{Monodromies of Neveu-Schwarz $B$ fields}\label{sec:UVB}

In this section we will illustrate the general considerations of \S\ref{sec:setupB} in a concrete framework for moduli stabilization.
We will exhibit a simple flattening effect \cite{flattening}\ in which the axion potential energy participates in the stabilization of a complex modulus $u$, whose adjustment reduces the power in the axion potential from a fiducial value $p_0=4$ to $p=3$, at fixed values of the other moduli.

Later, in \S\ref{iibRiem}, we will recover this effect within a class of string compactifications which also stabilize the volume and string coupling.  In short, the inflationary axion will arise from the NS-NS $B$ field in compactification of type IIB string theory on a product of Riemann surfaces, $\Sigma_1 \times \Sigma_2 \times \Sigma_3$, with moduli stabilized as in \cite{Saltman} by a combination of fluxes and $(p,q)$ 7-branes.

\subsection{Complex structure adjustment along a $B$ axion trajectory}\label{sec:fourtothree}

Consider type IIB string theory, but including the effective flux coupling T-dual to the term
$|F_0 B\wedge B|^2$ in the action (\ref{mzeroB}) for massive type IIA string theory.  As explained in
the previous section, in the presence of background fields (\ref{IIBfields}) \cite{massiveII}, we effectively have a term in the type IIB action of the form
\begin{equation}
S_{IIB} \supset \frac{1}{\alpha'^4} \int |F_1\wedge B\wedge B|^2 \, .
\end{equation}
from the $|\tilde F_5|^2$ term.  As usual in type IIB string theory, we must separately impose self-duality of $\tilde F_5$ \cite{Polchinski}.\footnote{The field configuration (\ref{IIBfields}) also contributes to the Chern-Simons term $\int C_4\wedge H_3\wedge F_3$ in the type IIB action, but in the flux and axion backgrounds considered below, the  relevant contribution $\int H_3\wedge F_1\wedge B$ will vanish.}

We begin by studying compactification on the  product of three two-tori, $(T^2)^3$, and later generalize to higher-genus Riemann surfaces.
For simplicity we will take the tori to be rectangular, with metric
\be\label{metricTtwo}
ds^2 = G_{mn}dy^m dy^n= \sum_{i=1}^3 L_1^2 (dy_1^{(i)})^2 + L_2^2 (dy_2^{(i)})^2\, ,
\ee
with $y_1\equiv y_1+\sqrt{\alpha^{\prime}}, y_2\equiv y_2+\sqrt{\alpha^{\prime}}$,
where $L_1$ and $L_2$ are dimensionless.  Denote $L^2=L_1L_2$, so the total internal volume $\vol$ is $L^6 {\alpha^{\prime}}^3$.
Introduce 3-form flux
\be\label{threeform}
F_3=(2\pi)^2\,\frac{Q_{31}}{\sqrt{\alpha'}}dy_1^{(1)}\wedge dy_1^{(2)}\wedge dy_1^{(3)}+(2\pi)^2\,\frac{Q_{32}}{\sqrt{\alpha'}}dy_2^{(1)}\wedge dy_2^{(2)}\wedge dy_2^{(3)}\, ,
\ee
where  the superscript labels the three two-tori and with
${Q_{31}, Q_{32}} \in \mathbb{Z}$.
That is, we have ${Q}_{31}$ units of flux on the product of the three $y_1^{(i)}$ cycles and  ${Q}_{32}$ units of flux on the product of the three $y_2^{(i)}$ cycles.

We now include quantized 1-form flux in the symmetric configuration
\be\label{oneform}
F_1=\frac{Q_1}{\sqrt{\alpha^{\prime}}} \sum_{i=1}^3 dy_1^{(i)}\, ,
\ee with $Q_1\in \mathbb{Z}$, so that $Q_1=\int dy_1^{(i)} F_1$.

The periods of $B$ on each individual $T^2$ give rise to the axions of primary interest:
\begin{equation}
b^{(i)} \equiv \frac{1}{\alpha^{\prime}}\int_{T^2_{(i)}}B\, ,
\end{equation}
so that\footnote{In our conventions, $b^{(i)}$ and $L$  are dimensionless, while $y_i$  have dimensions of length, and $B$  has the dimensions of length squared (so its components $B_{MN}$ are dimensionless, as are the components of the R-R potentials $C^{(p)}_{M_1\dots M_p}$).}
\be\label{Bs}
B=\sum_{i=1}^3 {b^{(i)}} dy_1^{(i)}\wedge dy_2^{(i)} + \ldots
\ee
where the ellipses indicate additional axions from periods of $B$ on two-cycles consisting of pairs of one-cycles from two distinct tori.\footnote{Axions involving distinct tori could be projected out  by a suitable orbifold action.}

We will first study the dynamics of the symmetric configuration
\be\label{bs}
b^{(1)}=b^{(2)}=b^{(3)} \equiv b \, ,
\ee
and will address the stability of the relative coordinates $b^{(i)}-b^{(j)}$, $i\neq j$, in \S\ref{transverse} below.
Upon dimensional reduction, the four-dimensional  Lagrange density for the scalars
takes the form (before converting to Einstein frame)
\be\label{action}
{\cal{L}}=\frac{a(t)^3}{\alpha'} \left\{\frac{L^6}{g_s^2}\left(\frac{\dot u}{u}\right)^2+\frac{L^6}{g_s^2}\left(\frac{\dot L}{L}\right)^2 +\frac{L^6}{g_s^2}\frac{\dot b^2}{L^4}-\frac{L^6}{\alpha'} \frac{Q_1^2}{L_1^2}\Biggl[\frac{b^4}{L^8}+\frac{b^2}{L^4}+1\Biggr]-\frac{L^6}{\alpha'}\left(\frac{Q_{31}^2}{L_1^6}+\frac{Q_{32}^2}{L_2^6}\right) \right\}\, ,
\ee
up to coefficients of order unity that we suppress.
Notice that the kinetic term for $b$ depends on $L$.  The dependence on $L$ in the various terms in (\ref{action}) is readily obtained from the metric (\ref{metricTtwo}), which  enters via the overall volume and the inverse metric components in the contractions $\tilde F_{\mu_1\dots\mu_n} G^{\mu_1\nu_1}\dots G^{\mu_n\nu_n} \tilde F_{\nu_1\dots\nu_n}$,  reflecting the dilution of the fluxes at large volume.

Stabilization of $L$ (and the remainder of the moduli) will be described in a particular class of examples in \S\ref{iibRiem} below, but it is useful first to examine the axion dynamics if $L$ is imagined to be fixed, as may happen in a variety of different ways in the string landscape.
The key phenomenon is that the energy built up in the $b^4$ term induces an adjustment of the complex structure modulus $u=L_2/L_1$, flattening the potential for $b$.

The combination in square brackets is
\be\label{sq}
\frac{b^4}{L^8}+\frac{b^2}{L^4}+1=\frac{\phi_b^4}{M_P^4}+\frac{\phi_b^2}{M_P^2}+1\approx  \frac{b^4}{L^8}\, ,
\ee
where $\phi_b$ is the canonically normalized inflaton, which satisfies $\phi_b \gg M_P$ in the regime of interest for inflation.  Correspondingly, we will drop the
constant
and quadratic terms in the square brackets from now on, in our analysis of the diagonal axion mode (\ref{bs}).

Using $L^2=L_1L_2$, writing $u=L_2/L_1$, and converting to Einstein frame gives the potential
\be\label{VE}
V\sim M_P^4\frac{g_s^4}{L^{12}}\left( \frac{Q_1^2}{L^4} u b^4+Q_{31}^2 u^3+\frac{Q_{32}^2}{u^3}\right)\, .
\ee
This potential stabilizes $u$, since it grows at large $u$ and at small $u$.  For simplicity of presentation
let us work in the regime where the second term in (\ref{VE}) can be neglected compared to the first and third terms.  Minimizing $u$ yields the $b$-dependent vev
\be\label{ub}
u\approx \frac{3^{1/4} L}{b}\sqrt{\frac{Q_{32}}{Q_1}} \propto \frac{1}{b}\, .
\ee
Substituting (\ref{ub}) into (\ref{VE}), and assuming that the kinetic energies are subleading in this dynamics, we see that the net effect is a flattening from $V\propto\phi^4$ to
\be\label{eqfourtothree}
V\propto \phi^3 \ .
\ee
In  the next two subsections we will verify that it is self-consistent to  neglect kinetic energies as a source of backreaction
(\S\ref{sec:kinetic}), and to restrict attention to the symmetric-combination axion field $b$ in (\ref{bs}), omitting the relative coordinates
$b^{(i)}-b^{(j)}$, $i\neq j$
(\S\ref{transverse}).

Granting  these facts,
and anticipating a full UV completion (to be discussed in \S\ref{iibRiem}), let us next estimate the scale of the parameters required for inflationary phenomenology, and check that the complex modulus $u$ is not driven to too extreme a value.  A detection of $r\approx 0.1$ corresponds to (roughly)
\be\label{parametersizes}
\frac{g_s^4}{L^9} \frac{4}{3^{3/4} }Q_1^{3/2} Q_{32}^{1/2} \sim 10^{-12}\, ,
\ee
where we folded in the super-Planckian regime $\phi\sim 10 M_P$ applicable to the early period of inflation.
This is straightforward to match with a  moderately weak string coupling $g_s\lesssim 1/10$ and moderately large compactification radius $L \gtrsim 10$, depending on the size of the fluxes $Q_1,Q_{32}$.

Finally, let us check that we do not require such an extreme value of $u=L_2/L_1$ that new degrees of freedom appear.  In particular, if
$L_2$ were too small ($L_2 \lesssim 1$, corresponding to a length $L_2\sqrt{\alpha^{\prime}}$ below the string scale), this would lead to light winding modes; we will avoid this regime.
In a more generic situation with a curved manifold, such as those we will study in \S\ref{iibRiem}, we require a
large radius of curvature.
Note that this does not in general require 1-cycle sizes to be larger than string scale; in the Riemann surface examples of \S\ref{iibRiem}, $u$ will be a complex structure modulus that does not change the curvature radius $\sim \sqrt{L_1L_2 \alpha^{\prime}}$.

The mass squared of the winding modes depends on the spin structure of the circle:  if fermions are anti-periodic, there is an unstable mode (winding tachyon)  for radii near the string scale,
whereas for a periodic spin structure there is no such instability.  In either case our model is safe from winding string effects, as follows.
Let us write (\ref{ub}) as
\be\label{ubcanonical}
u\sim \frac{L_2}{L_1} \sim  \frac{M_P}{\phi}\frac{1}{L}\sqrt{\frac{Q_{32}}{Q_1}} \gtrsim 10^{-1} \frac{1}{L}\sqrt{\frac{Q_{32}}{Q_1}}\, .
\ee
In (\ref{parametersizes}) we found $L=\sqrt{L_1L_2}\gtrsim 10$, with this inequality saturated for flux quantum numbers of order 1.  In that case,  (\ref{parametersizes}) and (\ref{ubcanonical}) would be satisfied by $L_1\sim 10^2, L_2\sim 1$.  This is already safe, and can be relaxed further (to larger $L_2$) using the fluxes $Q_1, Q_{32}$, allowing us to avoid extreme  values of the complex structure.

\subsection{Kinetic energies}\label{sec:kinetic}

That the kinetic  energies of the axions and moduli
are negligible here can be seen as follows.  First, the axion kinetic term depends on the size modulus $L$, which we have
temporarily
assumed to be stabilized independently.  Let us take the terms in the scalar potential that stabilize $L$ to be at least as large as the inflationary potential energy, and to be perturbative in $1/L$.  The explicit examples in the next section will satisfy this criterion.  The next step is to note that during inflation, the axion kinetic term is smaller than this inflationary potential by a factor of  $\varepsilon=\frac{\dot\phi^2}{H^2M_P^2}$.  So the axion kinetic  energy is a subleading source in the equation of motion for $L$.

It is likewise easy to show that the kinetic energy $u$ is also subleading in the dynamics, even though $u$ evolves during inflation, being yoked to the axion $b$ by (\ref{ub}).  We can write the $u$ degree of freedom in terms of the corresponding canonically normalized scalar field $\nu$:
\be\label{unu}
u\equiv e^{c_\nu\nu/M_P}\, ,
\ee
where $c_\nu$ is a constant of order 1.  The equation of motion for $\nu$ is
\be\label{nueom}
\ddot\nu + 3 H\dot\nu =-\partial_\nu V\, .
\ee
In (\ref{ub}) above, we approximated the solution to (\ref{nueom}) by setting the right hand side of  (\ref{nueom})  to zero, by balancing two individual terms in the potential against each other.
An individual term $V_{(i)}$ on the right hand side is of order
\be\label{nuV}
\partial_\nu V_{(i)}\sim \frac{V_{(i)}}{M_P}\sim H^2 M_P \sim \frac{V}{M_P}\, .
\ee
We will see shortly that each term on the left hand side of (\ref{nueom}) is much smaller than $V/M_P$, so that it was indeed a good approximation to solve for the dynamics of $u=e^{c_\nu \nu/M_P}$ by setting $\partial_\nu V=0$.

By differentiating the relation (\ref{ub}), we see that
\be\label{nudot}
\dot\nu \sim M_P\frac{\dot b}{b}\sim M_P\frac{\dot\phi}{\phi}\, ,
\ee while $\ddot\nu$ has terms of order $M_P \ddot\phi/\phi$ and $M_P\dot\phi^2/\phi^2$.  We have
\be\label{nudotH}
H\dot\nu \sim M_P H\frac{\dot\phi}{\phi}\sim \sqrt{\varepsilon}\frac{M_P}{\phi}\frac{V}{M_P} \ll \partial_\nu V_{(i)}\, ,
\ee
where in the last step we used (\ref{nuV}).  Similarly,
\be\label{nuddotineq}
M_P \frac{\ddot\phi}{\phi} ~~ \ll ~~ \frac{M_P}{\phi}\partial_\phi V \sim \left(\frac{M_P}{\phi}\right)^2\frac{V}{M_P}\sim  \left(\frac{M_P}{\phi}\right)^2 \partial_\nu V_{(i)} ~~ \ll ~~ \partial_\nu V_{(i)}\, ,
\ee
where we used the fact that in slow roll inflation, $\ddot\phi\ll\partial_\phi V\sim V/\phi \sim (M_P/\phi)(V/M_P)$.
Finally, the remaining contribution to $\ddot\nu$ is small:
\be\label{ddnutwo}
M_P\frac{\dot\phi^2}{\phi^2} \sim H^2 M_P\times \varepsilon \frac{M_P^2}{\phi^2} ~~ \ll ~~ \partial_\nu V_{(i)}\, .
\ee
In sum, the kinetic energies are all subleading in the dynamics.

\subsection{Transverse axion directions} \label{transverse}

Next, let us analyze the `relative' axion directions $b^{(i)}-b^{(j)}$, $i\neq j$, transverse to the configuration (\ref{bs}).  These transverse directions break a symmetry, and are guaranteed to lie at an extremum of the potential.  A positive mass squared in that direction, or a negative mass squared with $|m^2|\ll H^2$, does not represent an instability as we will see.

Let us focus on a given pair that contributes to $F_1\wedge B\wedge B$, say $b^{(1)}$ and $b^{(2)}$.
 Writing $b_\pm=b^{(1)}\pm b^{(2)}$ and similarly for the canonical fields $\phi_\pm$, the relevant terms in the potential are
\be\label{potpm}
V\sim
 \lambda(u) \left( \frac{\phi_+^2}{M_P^2}+\frac{\phi_-^2}{M_P^2}+\frac{\phi_+^4}{M_P^4}+\frac{\phi_-^4}{M_P^4}-\frac{\phi_-^2}{M_P^2}\frac{\phi_+^2}{M_P^2} \right).
\ee
Here we have allowed for dependence on $u$, which in the example just discussed led to a flattening to a cubic potential for $\phi_+$.
The last term in (\ref{potpm}) introduces a negative mass squared for $\phi_-$ that is of the same order as the positive mass squared of the perturbation $\delta\phi_+$.  This is well below the Hubble scale at large field values: for our power law potentials,
\be\label{masses}
|m_{\delta\phi}^2|\sim \partial_\phi^2 V\sim \frac{V}{\phi^2}\sim H^2 \frac{M_P^2}{\phi^2} ~~ \ll ~~ H^2.
\ee
There is  also a subleading positive contribution from the quadratic terms (which descend from the effective $|F_1 B|^2$ coupling).
Depending on the details of specific models, additional positive contributions can arise, for example from the $|F_3\wedge B|^2$ term, which depends on an independent set of flux quantum numbers.  In any case,  even before  taking into account any positive contributions, we obtain
parametrically mild instabilities, $|m_{-}^2|\ll H^2$, in the transverse directions.  Thus, while fluctuations of $\phi_-$ could contribute to the primordial perturbations,  instabilities in the $\phi_-$  direction do not prevent prolonged inflation.  Moreover, we do not need to sit precisely at $\phi_-=0$: inflation along the $\phi_+$ direction dominates even if we turn on $\phi_-$ as long as $\phi_-\ll\phi_+$ that ensures $|\partial_{\phi_+}V|\gg |\partial_{\phi_-}V|$.

\section{Embedding in Riemann Surface Compactifications} \label{iibRiem}

We now turn to embedding the preceding construction in a scenario for moduli stabilization.
Because the volume and string coupling will have finite masses in the stabilized vacuum, their vevs will be able to adjust to some degree, suggesting further flattening beyond that already evident in (\ref{eqfourtothree}).  This depends on the relative strengths of the terms in the potential that stabilize the various moduli, of which we will exhibit a few different cases.

\subsection{A concrete setup}

A natural class of compactifications to consider for this purpose is \cite{Saltman}, for two reasons.  First, the internal space (a product of Riemann surfaces) contains one-cycles that $F_1$ flux can thread.  Secondly, this mechanism for moduli stabilization (among others) comes equipped with relatively high potential barriers against runaways to weak coupling and large radius, a feature that fits naturally with the high energy scale of large-field inflation.

This will provide additional examples realizing the general mechanism of monodromy inflation.
As with previous realizations its role is to exhibit UV complete examples,  and in the present work an additional motivation is to explicitly map out a broader range of phenomenological predictions including the tensor to scalar ratio $r$.  Any given realization is not to be taken literally, since in the string landscape there are many arbitrary choices made in choosing a total dimensionality, a compactification manifold (or generalization), fluxes, defects, and other sources.  The mechanism itself --- the unwinding of the potential in the presence of generic branes and fluxes --- is rather robust; very specific realizations such as those developed here are meant simply as proofs of principle.

In order to incorporate axions from the Neveu-Schwarz $B$ field, we must check their compatibility with the ingredients involved in \cite{Saltman}.  In the latter mechanism for moduli stabilization, combinations of $(p,q)$ 7-branes triply intersect --- as in \cite{GKP}\cite{KKLT}\ --- to produce a source of negative tension scaling like that of orientifold 3-planes (O3).  The resulting negative term in the four-dimensional effective potential is useful for stabilizing the Riemann surface sizes and the string coupling \cite{Saltman}.  The coefficient of this term scales like $n_7^3$, where $n_7$ is the number of 7-branes, enabling it to compete with the positive terms in the potential, including that coming from the negative curvature of the Riemann surfaces (along with the 7-brane tensions).

\begin{figure}[t!]
\begin{center}
\includegraphics[width=5cm]{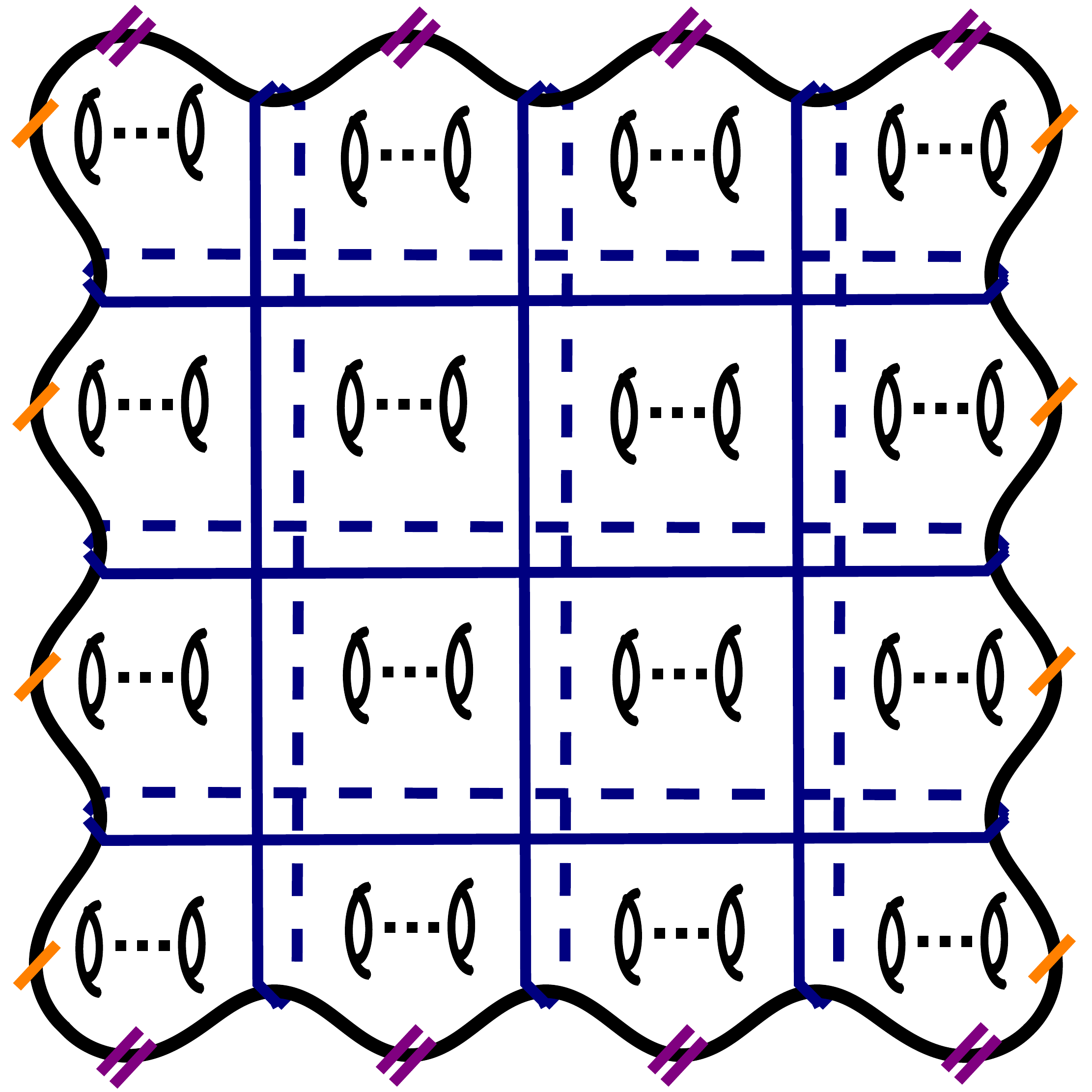}
\end{center}
\caption{An example of a very symmetric Riemann surface configuration, with the loci along which various sectors of 7-branes sit marked in blue.  As drawn, the 7-branes lie on contractible cycles, thereby
automatically satisfying Gauss's law constraints.   To create a Riemann surface with additional symmetry, we can impose periodic boundary conditions, cutting out holes where marked by single or double slashes and identifying them as indicated.  In that case each 7-brane at one location needs to be balanced by an antibrane elsewhere, a configuration also consistent with the setup in \cite{Saltman}.  The $F_1$ flux and legs of the $B$ field described in the text lie along the nontrivial $a$-cycles and $b$-cycles of the manifold.
When microscopic consistency conditions from the orientifold projection require  components of $B$ to vanish at the positions of the 7-branes, this can be achieved via suitable linear combinations as in (\ref{Bcycles}).}
\label{RSfig}
\end{figure}

\begin{figure}[htbp]
\begin{center}
\includegraphics[width=10cm]{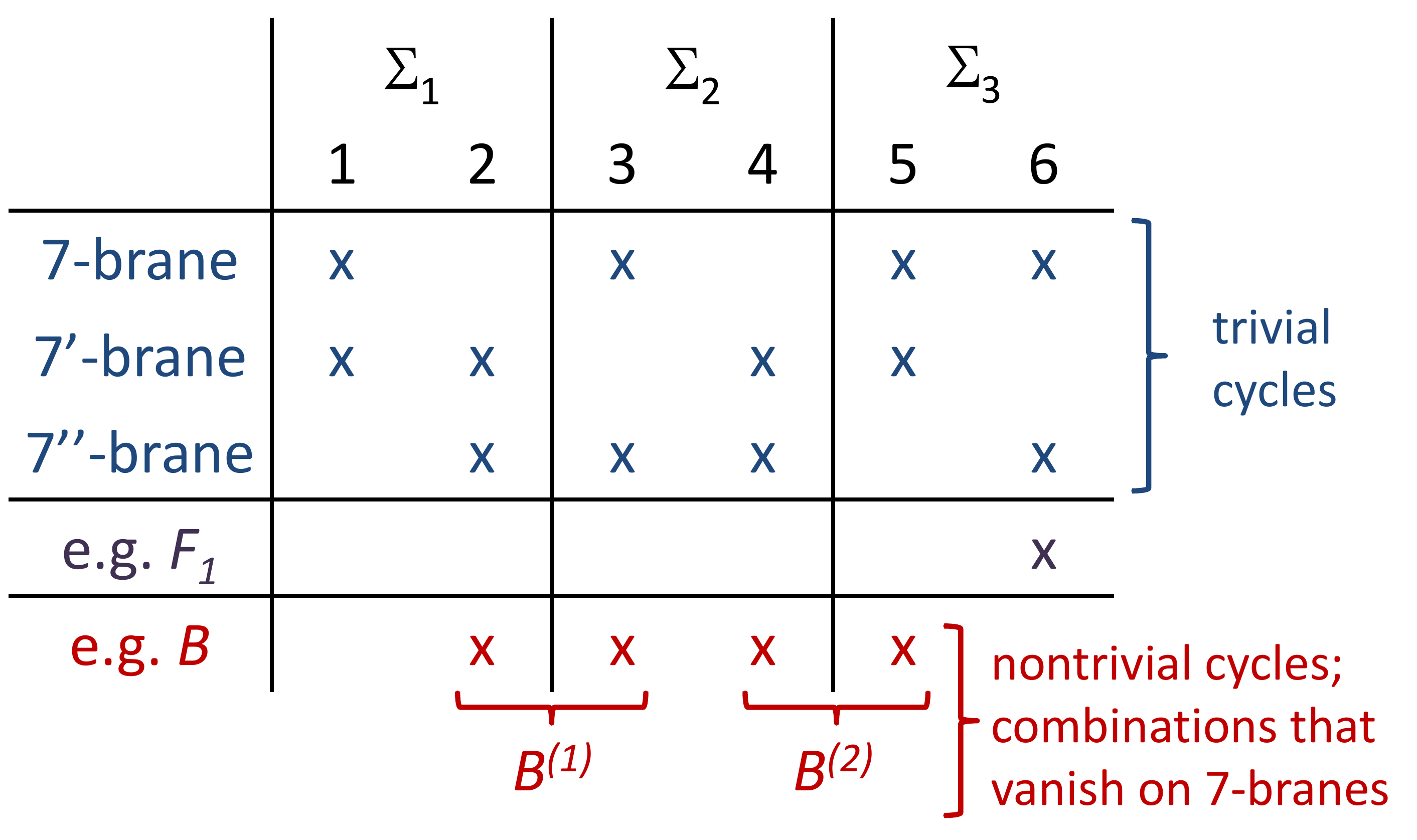}
\end{center}
\caption{The orientations of some of the ingredients.  The 7-branes lie on the blue cycles in figure \ref{RSfig}, while the $B$ field legs and $F_1$ lie on appropriate combinations of $a$- or $b$-cycles around the handles.  For example, the field $B^{(1)}$ has both legs parallel to the $7''$ sector of 7-branes, but $B^{(1)}$ vanishes at the position of the $7''$ branes if we take a linear combination with opposite orientations around the cycles on either side of each $7''$ brane on the Riemann surface $\Sigma_1$ depicted in figure \ref{RSfig}.}
\label{RStable}
\end{figure}

Let us simplify the construction \cite{Saltman} in the following way, preserving its essential features.  Wrap the 7-branes whose triple intersections give O3 tension on homologically trivial cycles --- the necks of the higher-genus Riemann surfaces as in figures \ref{RSfig}\ and \ref{RStable}\ below.  This automatically satisfies Gauss's law for all charges in the problem, and provides a symmetric, metastable configuration of these ingredients.

To be specific, as mentioned in \cite{Saltman}\ we may consider the combinations of 7-branes that behave outside their core like an O7-plane plus four D7-branes, the so-called SO(8) combination \cite{Senseven}.  However, we emphasize that unlike in the static, supersymmetric examples of that combination of 7-branes, in a system like ours with other forces at play, the 7-branes do not induce the asymptotic deficit angle of \cite{Senseven}.\footnote{In the case that we wrap these on contractible cycles, we can remove the mobile D7-branes by contracting them to a point.}

This point is worth elaboration: it is a string-theoretic version of the following standard physics.  Consider 2+1 dimensional gravity coupled to massive particles. In a static solution of the equations of motion, a massive particle induces a deficit angle proportional to its mass and the three-dimensional Newton constant $G_{N,3}$ \cite{threedGR}.  In these static solutions, the amount of matter is bounded by $m_{total}<1/ 2G_{N,3}$; when one saturates this the solution becomes compact.  But non-static solutions exist, including matter-dominated FRW expansion in the 2+1 dimensional theory.  In those solutions the amount of matter is not bounded:  it simply determines the rate of expansion via the Friedmann equation.  Similarly, 7-branes are not limited to the number --- namely, 24 --- that yields a static compact solution.  Examples of simple time-dependent solutions involving additional 7-branes include \cite{expandingF}.
In our application, the equations of motion have contributions from various sources --- the curvature, 7-branes and associated O3-planes, and fluxes.  The number of 7-branes is not constrained to be 24 in this more general context.

In our compactification, we can introduce $B$ fields along cycles as indicated in the figures.  In a static configuration, O7-planes would project out the constant mode of components of the $B$ fields with one leg parallel and one orthogonal to the orientifold.
However, modes of this $B_{\parallel\perp}$ field that vanish at the O7-plane are consistent: the orientifold  action essentially introduces a boundary condition that the $B$ field vanish  on the fixed locus.  We can satisfy this condition at the loci of the 7-branes by choosing suitable linear combinations of the $B$ fields, as explained in figure \ref{RStable}.

\subsection{Final powers: $p\approx 3, 2, 4/3, 2/3,  \dots$}\label{sec:BmonodromyVolumeFix}

Finally, we are in a position  to embed the example given in   \S\ref{sec:fourtothree}\ above into a compactification on a product of Riemann surfaces $\Sigma_1\times \Sigma_2\times \Sigma_3$.  In order to describe this, let us label the $a$-cycles of $\Sigma_1$ collectively as $1a$, the $b$-cycles of $\Sigma_1$ as $1b$, and so on, with associated one-forms $\omega_{1a}$, etc.

We will place $B$ fields and fluxes so as to generate a quartic term in the
$B$ axions
via the effective $|F_1\wedge B\wedge B|^2$ term.  The potential energy described in \cite{Saltman}\ depends on complex structure moduli analogous to $u$ in \S\ref{sec:fourtothree}, and on the volumes of the Riemann surfaces and the string coupling $g_s$.  Our analysis will
require generalizing equation (4.7) of \cite{Saltman}\ to include the potential energy in the axions from $B$, and keeping track of the dependence of the flux potentials $n_3^2, q_3^2, q_1^2$ on the relevant complex structure modulus $\tilde u$ analogous to $u$ in \S\ref{sec:fourtothree}.

We have a variety of choices for $B$ and  for $F_1$, as  well as for the orientations of $F_3, H_3$,  and $F_5$.
To begin, let us consider a simple configuration where we place $B$ fields along two-cycles of the form
\be\label{Bcycles}
B= b^{(1)} \sum_{I,J=1}^h\lambda_{IJ}\omega^I_{1b}\wedge\omega^J_{2a} + b^{(2)} \sum_{I,J=1}^h\lambda_{IJ}\omega^I_{2b}\wedge\omega^J_{3a} + b^{(3)}\sum_{I,J=1}^h\lambda_{IJ} \omega^I_{3b}\wedge\omega^J_{1a}\, .
\ee
This is analogous to the configuration \eqref{Bs} in \S\ref{sec:fourtothree}, but we orient the legs of the $B$ field as indicated so that they lie along 1-cycles on each Riemann surface factor, enabling us to enforce their vanishing at the positions of the 7-branes as described above (this is encoded in the signs $\lambda_{IJ}$).  Similarly we place $F_1$ flux along
\be\label{Fone}
F_1=Q_1(\omega_{1b}+\omega_{2b}+\omega_{3b})\, ,
\ee
where as we will discuss below, we either include $F_1$ on all such $a$- and $b$-cycles within each Riemann surface (a maximally symmetric choice) or instead thread  $F_1$ on a subset of these 1-cycles (or more generally, on the various cycles with different flux quantum numbers).
The combination of the $B$ field (\ref{Bcycles})  and the $F_1$ flux (\ref{Fone}) generates a contribution  to the potential energy from the effective $|F_1\wedge B\wedge B|^2$ term as in \S\ref{sec:fourtothree}.
In this configuration, the analogue of $u$, which in this section we are calling $\tilde u$, is $L_a/L_b$, where $L_a$ and $L_b$ are the sizes of the $a$-cycles and $b$-cycles of the Riemann surfaces.

We will find different behavior depending on whether the $F_5=dC_4$ flux required for the stabilization construction of \cite{Saltman} lies along different cycles from the $F_1\wedge B\wedge B$ contribution to the effective $\tilde F_5$, or if instead these fluxes overlap.  The latter case arises if we make the special, symmetric choice that all $h$ of the $a$- and $b$-cycles are threaded similarly by each type of flux.  The former case will provide a direct
embedding
of \S\ref{sec:fourtothree}.

At this point it is useful to introduce more of what we will need from the moduli stabilization mechanism of \cite{Saltman}.
Although the details are specific to the particular compactifications studied there, our analysis will expose some more general lessons.
The complex structure moduli of the Riemann surfaces are stabilized in \cite{Saltman}\ by a flux potential analogous to that in \cite{GKP},
arising from
the internal components of the generalized field strengths $\tilde F_p$ with Lagrangian
\begin{equation} \label{eqfppot}
V_{(p)} = \int \sqrt{-g} \tilde F_{\mu_1\dots\mu_p}g^{\mu_1\nu_1}\dots g^{\mu_p\nu_p}\tilde F_{\nu_1\dots\nu_p}\, .
\end{equation}
For each type of flux, (\ref{eqfppot}) reduces to a contribution to the four-dimensional effective potential that depends on complex structure moduli, flux quantum numbers, and axions as well as on the volume and dilaton.  Scaling out the latter dependencies, let us denote these flux potentials (as in \cite{Saltman}) as $H_3^2\to n_3^2$ and $\tilde F_p^2\to \tilde q_p^2$.  We will be interested in their dependence on complex structure moduli (including the analogue of $u$ in the simple model \eqref{VE} of \S\ref{sec:fourtothree}) as well as their dependence on the axions descending from $B$.

The string coupling $g_s$ and the volume $\vol$ of the product of Riemann surfaces are minimized by a potential ${\cal U}$ that realizes a combination of 2-term and 3-term perturbative stabilization mechanisms (cf.~e.g.~\cite{threetwo}). Including the generalized fluxes $\tilde F_p$, equation (4.5) of \cite{Saltman}\ becomes
\be\label{UnewI}
{\cal U} \sim  M_P^4\left\{\frac{h+n_7-1}{\sigma^2} -  \frac{N_7}{\sigma^3}+\frac{\tilde q_5^2}{\sigma^4}
+ \frac{n_3^2 }{\sigma^2\vol^{2/3}}+\tilde q_3^2  \frac{\vol^{2/3}}{\sigma^4}+q_1^2\frac{\vol^{4/3}}{\sigma^4} \right\}\, ,
\ee
where $\sigma\equiv g_s^{-1}\vol^{2/3}$, $h$ is the genus of each Riemann surface, and $n_7$ and $N_7\propto n_7^3$ are discrete parameters associated with the 7-branes in the construction \cite{Saltman}.  This potential metastabilizes $\sigma$ with a three-term structure; in \cite{Saltman}, the case in which the first three terms in (\ref{UnewI}) dominate over the others  was emphasized.  It is necessary for example that the three-form flux terms (i.e.~those proportional to $n_3^2$ and $\tilde q_3^2$) be at least marginally subdominant, since otherwise upon integrating out $\vol$ they produce a positive term scaling like $\sigma^{-3}$.  The stabilization of $\sigma$ requires the negative term in (\ref{UnewI}) to be sufficiently strong.    Moreover, in order for the $q_1^2$ term to at most marginally compete\footnote{The $q_1^2$ term need not be completely subdominant (the regime studied for simplicity in \cite{Saltman}).  It would consistent to let the $q_1^2$ term be large enough that it combined with the $n_3^2$ terms stabilizes $\vol$, leading to a positive term scaling like $\sigma^{-8/3}$ which combines with the second and third terms in (\ref{UnewI}) to stabilize $\sigma$.  In any case the $q_1^2$ term is at most marginally competitive with the leading terms, leading to (\ref{qonefive}).}  with the first three terms, we
must
have a hierarchy
\be\label{qonefive}
q_1^2 \le \frac{\tilde q_5^2}{\vol^{4/3}} \ll \tilde q_5^2\, .
\ee
In particular, when the axion $b$ goes to zero,
so that $\tilde{q_5} \to q_5$, we require $q_1\ll q_5$ at large volume $\vol$.
Given this, the flux terms $\sim n_3^2$, $\sim \tilde q_3^2$ in \eqref{UnewI}
stabilize the volume $\vol$ with an essentially two-term structure,  diverging at small or large $\vol$ (for fixed $\sigma$).  The complex structure moduli are stabilized by fluxes via a similar two-term structure encoded in the flux potential \cite{Saltman}\cite{GKP}.   Intuitively, at fixed volume and string coupling, fluxes on dual $a$-cycles and $b$-cycles cost the system increasing energy if the relative sizes of these cycles change in either direction, as then the flux becomes more concentrated.

Let us for simplicity consider 3-form fluxes that have vanishing wedge product with the $B$ fields (\ref{Bcycles}).
This is the case for the simplest generalization of \eqref{VE} in \S\ref{sec:fourtothree}\ to Riemann surfaces, with the three-form fluxes threading cycles consisting of products of $a$-cycles or products of $b$-cycles, as in (\ref{threeform}).    Then we will have two cases of interest, depending on whether $(F_1\wedge B\wedge B) \wedge dC_4$ is nonzero; this gives two different behaviors for the axion-dependence in $\tilde q_5^2$. Somewhat schematically,
\bea\label{tildefive}
\tilde q_5^2 &\sim& q_5^2 + 2 q_5 q_1 b^2 + q_1^2 b^4 ~~~~~ {\rm overlapping ~ case ~ (i)}\, , \nonumber   \\
\tilde q_5^2 &\sim& q_5^2 + q_1^2 b^4 ~~~~~~~~~ {\rm non-overlapping ~ case ~ (ii)}\, ,  \\ \nonumber
\eea
where these terms depend implicitly on the complex structure moduli in a way that depends on the 1-cycles they thread (as we will make explicit below in specific examples).
The canonical field is\footnote{This expression  omits  possible dependence on the genus $h$.}
$\phi\sim b M_P/L^2 \sim b M_P/\vol^{1/3}$ (for curvature radius $L\sqrt{\alpha'}$), but it is useful at least at first to analyze the action in terms of $b$, while making sure to treat the $\vol$-dependence in the kinetic term consistently.

Before moving to the complete models realizing the complex structure flattening mechanism in \S\ref{sec:fourtothree}, which will arise in case (ii) of (\ref{tildefive}), let us begin with case (i).
This arises from the most symmetric choice we can make, with the fluxes threading cycles in all the handles of the Riemann surface democratically.
For this first class of examples, we will keep the complex structure moduli stabilized as in \cite{Saltman}, and focus on the dependence on $g_s$ and $\vol$ (equivalently, the dependence on $\sigma$ and $\vol$).
To implement this, we can add the analogue of (\ref{Bcycles}) and (\ref{Fone}) in which we exchange the $a$- and $b$-cycles, and also consider an arrangement of $F_3, H_3,$ and $F_5$ that is symmetric under this exchange.  This stabilizes the ratios of $a$- and $b$-cycle sizes (the complex structure moduli) at a value of $\tilde u\sim L_a/L_b$ of order 1.  (The symmetry is not essential here; more generally one can just choose $F_1$ and the other fluxes so that the inflationary potential depending on $b$ and the pure flux terms agree on the minimum in the $\tilde u$ direction.)

Given this, the volume, dilaton, and axion dynamics works as follows for case (i).  We may first minimize $\sigma$ at its minimum $\sigma_{min}$ determined by the (dominant) first three terms in (\ref{UnewI}), with $\tilde q_5\approx q_5$.  Then
the final step of volume and dilaton stabilization, in the presence of the axion (\ref{Bcycles}), simply requires generalizing equation (4.7) from \cite{Saltman}\ in the following way: defining
\be\label{Ch}
C_h\equiv \frac{h-n_7-1}{N_7},
\ee
and incorporating the quartic term in the axion field $b=b^{(i)}, i=1,2,3$,
we obtain
\be\label{UnewIIi}
{\cal U}|_{\sigma=\sigma_{min}} \sim M_P^4\left\{ C_h^2 n_3^2 \frac{1}{\vol^{2/3}}+C_h^4 [q_3^2+q_1^2b^2] \vol^{2/3} + C_h^4 (q_5^2 + 2 q_5 q_1 b^2 + q_1^2 b^4)\right\}  ~~~~~ {\rm case~ (i)}\, ,
\ee
valid as long as $2 q_5 q_1 b^2 + q_1^2 b^4\le {\cal O}(q_5^2)$.  The first two terms here stabilize the volume $\vol$.

First, let us consider the case where $q_3^2 \gg q_1^2 b^2$.  Since in the moduli stabilization mechanism \cite{Saltman}\ we have $q_1\ll q_5$, cf.~(\ref{qonefive}), there is a window in which the inflationary potential is quadratic plus quartic in the axion, over many underlying periods of $b$.  A super-Planckian vev for the canonically normalized field  $\phi_b\sim \sqrt{h} M_P b/\vol^{1/3}$ (with $h$ the genus of the Riemann surfaces), keeping the axion terms at most marginally competitive with the $q_5^2$ term in (\ref{UnewIIi}), requires
\be\label{hineq}
 \frac{q_1 \vol^{2/3}}{h q_5}\le \frac{M_P^2}{\phi^2},
\ee
which is satisfied given a moderately large $h$ (as is required to be able to tune the cosmological constant in \cite{Saltman}) or given a significant hierarchy between $q_1$ and $q_5$.  We impose (\ref{hineq}) because if the axion terms dominated over the $q_5^2$ term, then the third term in (\ref{UnewI}) would vary over several orders of magnitude as $\phi_b$ varies from $\sim 10 M_P$ to $\sim M_P$, which would invalidate the three-term stabilization of $\sigma$.
The condition (\ref{hineq}) in turn ensures that the quadratic term in the round brackets of \eqref{UnewIIi} dominates over the quartic term in the axion, although the latter could become  significant at the outer edge of the super-Planckian regime.  (This latter effect may be interesting in light of the hints of a tension between Planck and BICEP2 \cite{postBICEP}, although that tension is {\it{highly}} uncertain given \cite{spergelWMAP}\cite{notension}\ as well as foreground unknowns.)

Since we have expressed the action in terms of $b$ rather than the canonical field $\phi\sim \sqrt{h} M_P b/\vol^{1/3}$, we must take into account the $\vol$ dependence in the kinetic term for $b$,  ${\cal S}_{kin}\sim \int M_P^2 h\dot b^2/\vol^{2/3}$.  However, this is easily subdominant in the dynamics of $\vol$, since the inflaton kinetic energy is much less than its potential energy (by a factor of the inflationary slow roll parameter $\varepsilon$), and this in turn may be easily kept smaller than each of the first two terms in
(\ref{UnewIIi}).  This last statement follows from the subdominance of the $q_1$ contribution to the moduli potential in \cite{Saltman}.  Altogether, from case (i) of (\ref{tildefive}) we have obtained a quadratic inflaton potential, crossing over to quartic at the boundary of its large field range (as it reaches the regime where the inflaton potential would destabilize the modulus $\sigma$).

In more general circumstances, as we will see momentarily in a specific example, the kinetic term can play a more nontrivial role in the dynamics; if $\vol$ depends on $b$, this affects the definition of the canonically normalized inflaton field $\phi_b$, which can  alter the ultimate power of the potential \cite{flattening}.  In the special case that the kinetic term after volume stabilization is proportional to $(\dot b/b)^2$, this  change can prevent inflation, as it renders a potential that is power law in $b$ exponential in terms of the canonically normalized field.  Without a sufficiently small coefficient in the exponent (which may arise in some cases, but not generally), or a separation of mass scales, this will not inflate.

Next, let us consider the case where $q_1^2 b^2 \gg q_3^2$ in the square brackets in
(\ref{UnewIIi}).  In this case, we obtain $b\propto 1/\vol^{2/3}$ from the first two terms in the potential (\ref{UnewIIi}).  This has two effects:  it introduces a linear term in $b$ in the potential, and it also changes the relation between $b$ and the canonically normalized inflaton field $\phi_b$ because of the $\vol$ dependence in the $b$ kinetic term
\be\label{kincan}
\frac{\dot b^2}{\vol^{2/3}}\propto \dot b^2 b \Rightarrow \phi_b\propto b^{3/2}
\ee
 (see \cite{flattening}\ for previous examples of this effect).
Given (\ref{kincan}), for the regime (\ref{hineq}) in which the quadratic term in $b$ $\propto q_1q_5 b^2$ dominates in the potential, one finds $p=(2/3)\times 2=4/3$:
\be\label{fourthirds}
V(\phi_b)\approx \mu^{8/3}\phi^{4/3}\, .
\ee

Before moving on to complex structure adjustments, we can obtain another class of models  from the case $p_0=2$.  Consider a set of $B$ fields for which $B\wedge B$ vanishes (or is negligible), obtainable by appropriate distribution of the legs of $B$ among the handles of the Riemann surfaces.  As above, we take the dominant flux terms --- including $q_5$ --- to stabilize the corresponding complex structure moduli, as in \cite{Saltman}. The field configuration (\ref{IIBfields}) gives a contribution to $F_3$ of the form $F_1\wedge B$, orthogonal to the components of $F_3$ we prescribed above (in which the three legs are either all on $a$-cycles or all on $b$-cycles).   In the absence of  $B\wedge B$ contributions, the  potential takes the form
\be\label{UnewIIiagain}
{\cal U}|_{\sigma=\sigma_{min}} \sim M_P^4\left\{ C_h^2 n_3^2 \frac{1}{\vol^{2/3}}+C_h^4 (q_3^2+q_1^2 b^2) \vol^{2/3} + C_h^4 q_5^2 \right\}  ~~~~~ {\rm case~ (i)}\, .
\ee
The result is a linear contribution to the potential in $b$, as in the previous example, but here there are no additional quadratic or quartic terms.  The kinetic term works as in (\ref{kincan}), giving $p=2/3$:
\be\label{twothirds}
V\approx \mu^{10/3} \phi^{2/3}\, .
\ee

Let us next move to case (ii) of (\ref{tildefive}), which gives an a priori quartic dependence on $b$, i.e. a fiducial power $p_0=4$.
In these next examples, we will also incorporate a more general complex structure dependence, including dependence on a modulus $\tilde u=L_a/L_b$ describing the ratio of $a$- and $b$-cycle sizes in some subset of Riemann surface handles.
\be\label{UnewIIii}
{\cal U}|_{\sigma=\sigma_{min}} \sim M_P^4\left\{ C_h^2 n_3(\tilde u)^2 \frac{1}{\vol^{2/3}}+C_h^4 q_3(\tilde u)^2 \vol^{2/3} + C_h^4 [q_5(\tilde u)^2 + q_1(\tilde u)^2 b^4]\right\}  ~~~~~ {\rm case~ (ii)}\, .
\ee
The $\tilde u$ dependence arises from the dependence of the flux energies on the complex structure derived explicitly in \cite{Saltman}, along with the analogous complex structure dependence in the axion potential terms (coming from the $B$ dependence in the generalized fluxes $\tilde F_p$).  Depending on how we distribute the legs of the fluxes, each type of flux that threads 1-cycles within the Riemann surfaces can depend on $\tilde u\sim L_b/L_a$ as a combination of terms of order $\tilde u^{\pm 1}$, and for the three-forms $H_3$ and $F_3$ we can also have terms of order $\tilde u^{\pm 3}$ from fluxes threading a one-cycle of each of the three Riemann surfaces  (as in the model \eqref{VE} in \S\ref{sec:fourtothree}).  Our two-form potential $B$ threads two-cycles composed of a product of $a$- and $b$-cycles (\ref{Bcycles}), and so $b$ does not have any implicit $\tilde u$ dependence.

To obtain more general examples, we can break some of the symmetry assumed in the first set of examples described above.  There are two ways in which we can generalize:  (I) break the symmetry among the different pairs of $a$- and $b$-cycles on each Riemann surface, and/or (II) break the symmetry among the three Riemann surfaces.  We will next consider two sets of examples, in the first case relaxing the symmetry just in sense (I) and in the second set generalizing in the direction of both (I) and (II) together.  This will give us powers $p\approx 3$ and $p\approx 2$ respectively, starting from the fiducial power $p_0=4$.

First, let us consider particular subsets of pairs of $a$-cycles and $b$-cycles on which to thread the $F_1$ flux,  treating the handles of each Riemann surface less symmetrically.  In this case, the inflaton potential term $V\propto q_1^2 b^{p_0}$ with $p_0=4$ has a distinct dependence on the corresponding complex structure moduli $\tilde u_I=L_{aI}/L_{bI}$ (where the index $I=1,\dots, n_1$ runs over the handles threaded by $F_1$ --- taking at least the minimal number
required to respect the consistency conditions from the 7-branes).  Let us also separate the fluxes that stabilize the complex structure moduli in \cite{Saltman}\ into those with legs on these $n_1$ cycles (which we can label $\Delta F_p$) and those without legs on them (which we will call $F_p^{(0)}$).  The latter we can take to dominate in stabilizing the string coupling and volume in \eqref{UnewI}, which proceeds as described in \cite{Saltman}.  (There $F_1^{(0)}$ is not required, and we can consider for simplicity $F_1=\Delta F_1$, i.e.~only threading $F_1$ on $n_1$ of the cycles as just prescribed.)

The set of fluxes threading cycles within the $n_1$ handles, i.e.~the $\Delta F_p$ fluxes, includes some that depend on the combination $L_{aI}L_{bI}$, and others that depend on the ratio $\tilde u_I=L_{aI}/L_{bI}$.  The former combined with the $F_P^{(0)}$ fluxes stabilize the product $L_{aI}L_{bI}$ as in \cite{Saltman}, provided that one chooses large enough flux quantum numbers in these sectors so that this is a leading effect.

Finally, we can  address the  stabilization of $\tilde u_I=L_{aI}/L_{bI}$.
The $B$ fields and $F_1$, and the remaining $\Delta F_p$ fluxes stabilize this just as in the model \eqref{oneform}, \eqref{Bs}, and \eqref{VE} explained in \S\ref{sec:fourtothree}\ (replacing $u$ in that toroidal toy model with $\tilde u$ in the Riemann surface compactification).  The $b$ kinetic term depends only on the volume of a given handle (the product $L_{aI}L_{bI}$), not on $\tilde u\sim L_{aI}/L_{bI}$.  At the minimum in the $\sigma$ and $\vol$ directions, the $\tilde u$ dependence in the potential is of the form (cf (\ref{VE}))
\ba\label{Utildeu}
{\cal U}|_{*}\sim  M_P^4 C_h^2\left\{\left( \frac{\Delta n_{31}^2}{\vol_*^{2/3}}+\Delta q_{31}^2\vol_*^{2/3}\right)\tilde u^3 +\left( \frac{\Delta n_{31}^2}{\vol_*^{2/3}}+\Delta q_{31}^2\vol_*^{2/3}\right)\frac{1}{\tilde u^3}
+ C_h^2 q_1^2 b^4 \tilde u   \right\}\, ,
\ea
where ${\cal U}|_{*}$ is shorthand for ${\cal U}|_{\sigma=\sigma_*, \vol=\vol_*}$, and we stress that (\ref{Utildeu}) applies in case (ii).

As in the previous example, we work in the regime where the axion kinetic term is a subdominant source in the equation of motion for $L_{aI}L_{bI}$, leaving $L_{aI}L_{bI}$ stabilized as in \cite{Saltman}.  The kinetic term for $\tilde u$ is also subleading in the dynamics, as explained above in \S\ref{sec:kinetic}.
Altogether, stabilizing $\tilde u$ during inflation using the last two terms in (\ref{Utildeu}), as in \S\ref{sec:fourtothree}\ this gives a flattening to a cubic potential from the fiducial quartic potential,
\be\label{pthreeUV}
V \approx \mu\phi^3, ~~~ p_0=4\to p=3.
\ee

For another class of examples, we can relax the symmetry further and allow the three Riemann surface factors to behave differently.  Then, instead of a cubic dependence on $\tilde u$ in the three-form flux terms, we obtain $\tilde u^{\pm 1}$.  This, combined with the dependence $b^4\tilde u$ in the axion term, leads to $\tilde u\sim 1/b^2$, and $V\propto b^2$.  That is, this last class of examples produces to good approximation $\frac{1}{2}m^2\phi^2$ inflation,
\be\label{ptwoUV}
V \approx \frac{1}{2} m^2\phi^2, ~~~p_0=4\to  p=2.
\ee

It is clear from the examples  considered thus far that various powers appear, giving a wide range of (discretely different) values of $r$.
A quadratic potential is among them, coming either as the result of rigid stabilization with a quadratic potential, or via flattening from a quartic potential.  However, from the top down the quadratic model is not particularly special.  It is a classic model from the bottom up \cite{Andreichaotic}, and is simple in some sense.   But this simplicity may be illusory --- the field theory model alone does not account for quantum gravity effects (or particle physics or the cosmological constant).  From the top down, the monodromy mechanism for large fields that underlies this and other examples appears to be quite simple, with moduli stabilization introducing what complications there are in the problem.
As we have seen here, the inflationary dynamics itself can participate in a rather simple way in moduli stabilization, simplifying the latter somewhat.

\section{Monodromies of Complex Structure Moduli}\label{sec:UVII}

For special classes of compactification manifolds, such as Calabi-Yau spaces and Riemann surfaces, the metric deformations include complex structure moduli.  Their monodromies
play an important role in the mathematical structure of
these compactifications, particularly in the Calabi-Yau case where much of the moduli space geometry has been mapped out.  As mentioned above, these generalize the $\tau\to\tau+1$ symmetry for a torus, for which $Re(\tau)$ plays the role of an axion.

It is natural to consider these moduli as candidate inflatons, a topic we leave mainly for future work.
But as a start, it is straightforward to derive a close analogue of  the examples  given in \S\ref{sec:fourtothree}, starting from the T-duality between $B$ fields and angular metric deformations.
Specifically, we
T-dualize on the $y_2$ directions of the $T^2$ factors in that toy model.  For each torus, T-duality exchanges
\be\label{rhotau}
\rho = b+i \sqrt{G} ~~ \leftrightarrow ~~ \tau=\frac{G_{12}}{G_{22}}+i\frac{\sqrt{G}}{G_{22}} \equiv \tau_1+i\tau_2\, ,
\ee
where the metric of the $T^2$ is $ds^2=G_{MN}dy^Mdy^N$ and $\sqrt{G}= L^2$ is the volume.  The quasiperiodic direction under $b\to b+1$ maps to $\tau\to\tau+1$.  The effective flux coupling $|F_1\wedge B\wedge B|^2$ yielded the monodromy-induced quartic coupling in (\ref{VE}). Under the T-duality (\ref{rhotau}), $F_1$ dualizes to four-form flux $F_4$, and
and the three-form fluxes dualize to $F_0$ and $F_6$.  The resulting effective potential on the T-dual side is, for $\tau_1 \gg 1$,
\be\label{rhopot}
V\sim M_P^4 \frac{g_s^4}{L^{12}}\left( Q_4^2 \frac{\tau_1^4}{\rho_2 \tau_2^2}+\rho_2^3 Q_0^2 +\frac{Q_6^2}{\rho_2^3}\right)\, ,
\ee
as can be computed directly using the T-dual fluxes and metric, or by applying (\ref{rhotau}) to (\ref{VE}).  In parallel to the previous case, solving for $\rho_2$ here gives a cubic inflationary potential along the $\tau_1$ direction.
We leave  the study of generalizations that are not directly T-dual to previous examples as an interesting problem for the future.

\section{Conclusions}\label{sec:conclusions}

Monodromies  of axion fields are ubiquitous in string compactifications with sufficiently general fluxes or brane configurations. In this work we first provided an overview of the monodromy mechanism, emphasizing the genericity of the large field ranges induced by flux couplings along axion directions, as well as the role of the underlying discrete shift symmetry in protecting other aspects of the physics.   Just as the potential exhibits a branch structure with an underlying periodicity, as in figure \ref{strawmanargument}, there is a periodicity in the spectrum of branes wrapping the cycles that yield axions from higher-dimensional potential fields.  Inflation proceeds on one branch of the monodromy-extended potential, while these sectors of the spectrum remain periodic.

While it is straightforward to identify  compactifications containing fields and couplings  that appear suitable for large-field inflation,  stabilizing moduli remains  the primary technical complication, both in monodromy constructions and in all other scenarios for inflation in string theory.  The most detailed and explicit scenario presented  in this work  builds on the  construction of \cite{Saltman}, in which  the moduli of type IIB  string theory compactified on a product of Riemann surfaces are stabilized by  fluxes and $(p,q)$ 7-branes.   In this setting, the inflaton corresponds to an axion descending from  the NS-NS  two-form $B$, and the monodromy is a consequence of the coupling $|F_1\wedge B\wedge B|^2$ T-dual to the coupling $|m_0 B\wedge B|^2$  in massive type IIA  string theory.  We also related this via T-duality to monodromies in complex structure moduli space, which may provide another rich set of examples to explore;
constructing more explicit  and general examples  along those lines is an important task for the future.

We presented several new models of large-field inflation from axion monodromy, involving monomial potentials $\mu^{4-p}\phi^{p}$.
A key phenomenon
is {\it{flattening}} \cite{flattening}, in which the inflationary potential energy density makes a leading contribution to the potential for some of the moduli, whose vevs then adjust during the course of inflation, reducing the total energy.   The result is that an  exponent $p_0$ computed in the absence of flattening is reduced to $p<p_0$  by the  dynamical adjustment of the moduli.  In this work,  we exhibited examples with $p=3,2,4/3$, and $2/3$, realizing a large range of phenomenological predictions for the tensor to scalar ratio.   This includes a class of examples with
flattening from $p_0=4$ to $p=3$, somewhat analogous to the flattening from $p_0=2$ to $p=1$ in \cite{monodromyII}\cite{flattening}.   It would be extremely interesting to build from this experience to more systematically characterize the powers arising in monodromy inflation.   The present work, as well as \cite{gurari}\cite{flattening}, provide a modest start to this program, by incorporating the natural interplay between inflation and moduli stabilization.

The monodromy structure of string theory axions, and their duals among complex structure moduli and brane positions, has played an interesting mathematical role in the theory, and naturally generates large-field inflation.
Phenomenologically, the discrete examples of $p$ obtained in this and other works, and a more systematic generalization if that can be accomplished, relate directly to various cosmological observables.  It is of great interest to understand the spectrum of UV-complete values of $r$ (a detectable amplitude of tensor fluctuations being the main model-independent signature of monodromy inflation) as well as $n_s$ (which depends on $p$ and also on the number of fields involved \cite{recentmonodromy}).  In addition, one would like to map out the more detailed, but model-dependent signatures from the residual oscillations in the potential (\ref{Vform}) \cite{monodromyII}\ generated by the sectors of the physics that respect the underlying periodicity $\phi \to \phi+2\pi f$.  The search for such oscillations --- which has so far led to constraints \cite{oscillations}\ --- may be affected by the theoretical spectrum of possible values of $p$, and by the possibility of dynamical relaxation of the period $f$ and of the model-dependent amplitude $\Lambda^4$ of the oscillations during inflation.  All this provides ample motivation for further study.

\section*{Acknowledgements}

We thank X. Dong, S. Hartnoll, S. Kachru, R. Kallosh, N. Kaloper,  A. Lawrence, A. Linde, J. Maldacena, P. McGuirk, J. Polchinski, L. Senatore, S. Shenker, G. Torroba, E. Witten, and M. Zaldarriaga for useful discussions.
The research of L.~M.~was supported by NSF grant PHY-0757868.   The work of E.S. was supported  in part by the National Science Foundation
under grant PHY-0756174 and NSF PHY11-25915 and by the Department of Energy under
contract DE-AC03-76SF00515.  The research of  A.W. was supported by the Impuls und Vernetzungsfond of the Helmholtz Association of German Research Centres under grant HZ-NG-603.  A.W. would like to thank SITP,  where the main part of this work was completed, for their extraordinarily warm hospitality.   T.W. was supported by a Research Fellowship (Grant number WR 166/1-1) of the German Research Foundation (DFG).

\end{document}